\documentclass[12pt]{iopart}
\usepackage{epsfig,wrapfig}  

\def\la{\mathrel{\mathchoice {\vcenter{\offinterlineskip\halign{\hfil
$\displaystyle##$\hfil\cr<\cr\sim\cr}}}
{\vcenter{\offinterlineskip\halign{\hfil$\textstyle##$\hfil\cr<\cr\sim\cr}}}
{\vcenter{\offinterlineskip\halign{\hfil$\scriptstyle##$\hfil\cr<\cr\sim\cr}}}
{\vcenter{\offinterlineskip\halign{\hfil$\scriptscriptstyle##$\hfil\cr<\cr
\sim\cr}}}}}

\def\ga{\mathrel{\mathchoice {\vcenter{\offinterlineskip\halign{\hfil
$\displaystyle##$\hfil\cr>\cr\sim\cr}}}
{\vcenter{\offinterlineskip\halign{\hfil$\textstyle##$\hfil\cr>\cr\sim\cr}}}
{\vcenter{\offinterlineskip\halign{\hfil$\scriptstyle##$\hfil\cr>\cr\sim\cr}}}
{\vcenter{\offinterlineskip\halign{\hfil$\scriptscriptstyle##$\hfil\cr>\cr
\sim\cr}}}}}

\newcommand{\cray} {{\sc cr}}
\newcommand{\eas} {{\sc eas}}
\newcommand{\xm} {\ensuremath{X_{max}}}

\begin{document}

\title[Reconstructing Energy and Mass of Primary Cosmic Ray
Particles]{Methods of Determination of the\\
Energy and Mass of Primary Cosmic Ray Particles\\
at Extensive Air Shower Energies}

\author{Karl-Heinz Kampert\footnote[3]{e-mail: {\tt kampert@ik1.fzk.de}}}

\address{Universit\"at Karlsruhe (TH), Institut f\"ur Experimentelle 
Kernphysik}

\address{Forschungszentrum Karlsruhe, Institut f\"ur Kernphysik\\
P.O.B. 3640, D-76021 Karlsruhe, Germany}

\begin{abstract}
Measurements of cosmic ray particles at energies above $E \ga
5\cdot 10^{14}$ eV are performed by large area ground based air shower
experiments.  Only they provide the collection power required for
obtaining sufficient statistics at the low flux levels involved. 
In this review we briefly outline the physics and astrophysics
interests of such measurements and discuss in more detail various
experimental techniques applied for reconstructing the energy
and mass of the primary particles.  These include surface arrays
of particle detectors as well as observations of Cherenkov-
and of fluorescence light.  A large variety of air shower
observables is then reconstructed from such data and used to
infer the properties of the primary particles via comparisons to
air shower simulations.  Advantages, limitations, and
systematic uncertainties of different approaches will be
critically discussed.
\end{abstract}




\section{Introduction}

The cosmic ray (\cray) energy spectrum extends from about 1 GeV
to above $10^{20}$~eV. Over this wide range of energies the
intensity drops by more than 30 orders of magnitude.  Despite the
enormous dynamic range covered, the spectrum appears rather
structureless and can be well approximated by broken power-laws
$dN/dE \propto E^{-\gamma}$.  Up to energies of some $10^{14}$~eV
the flux of particles is sufficiently high allowing measurements 
of their elemental distributions by high flying balloon or
satellite experiments.  Such studies have provided important
implications for the origin and transport properties of \cray's
in the interstellar medium.  Two prominent examples are ratios of
secondary to primary elements, such as the B/C-ratio, which are
used to extract the average amount of matter \cray-particles have
traversed from their sources to the solar system, or are
radioactive isotopes, e.g.\ $^{10}$Be or $^{26}$Al, which carry
information about the average `age' of cosmic rays.  With many
new complex experiments taking data or starting up in the near
future and with a possibly new generation of balloons, this is a
vital field of research and has been subject to separate talks
presented at this Symposium.

Above a few times $10^{15}$~eV the flux drops to only one
particle per m$^{2}$ per year.  This excludes any type of `direct
observation' even in the near future, at least if high statistics is
required.  Ironically, one of the most prominent features of the
\cray\ energy spectrum is the steepening of the slope from
$\gamma \cong 2.7$ to $\gamma \cong 3.1$ at an energy just above
some $10^{15}$~eV. This is known as the `knee' of the spectrum. 
It was first deduced from observations of the shower size
spectrum made by Kulikov and Khristianson \etal in 1956
\cite{kulikov56} but it still remains unclear as to what is the
cause of this spectral steepening.  At an energy above
$10^{18}$~eV the spectrum flattens again at what is called the
`ankle'.  Data currently exist, though with very poor statistics,
up to $3 \cdot 10^{20}$~eV and there seems to be no end to the
energy spectrum \cite{agasa00,nagano00b}.

The origin and acceleration mechanism of these, so called, ultra-
and extremely high energy cosmic rays have been subject to debate
for several decades.  Mainly for reasons of the required power
the dominant acceleration sites are generally believed to be
shocks associated with supernova remnants.  Naturally, this leads to
a power law spectrum as is observed experimentally.  Detailed
examination suggests that this process is limited to $E/Z \la
10^{15}$\,eV. Curiously, this coincides well with the knee at
$E_{\rm knee} \cong 4 \cdot 10^{15}$\,eV, indicating that the
feature may be related to the upper limit of acceleration.  The
underlying picture of particle acceleration in magnetic field
irregularities in the vicinity of strong shocks suggests the
maximum energies of different elements to scale with their
rigidity $R=pc/Ze$.  This naturally would lead to an
overabundance of heavy elements above the knee, a prediction to
be proven by experiments.  A change in the \cray\ propagation with
decreasing galactic containment at higher energies has also been
considered.  This rising leakage results in a steepening of the
\cray\ energy spectrum and again would lead to a similar scaling
with the rigidity of particles but would in addition predict
anisotropies in the arrival direction of \cray's with respect to
the galactic plane.  Besides such kind of `conventional' source
and propagation models \cite{drury94b,berezhko99} several other
hypotheses have been discussed in the recent literature.  These
include the astrophysically motivated single source model of
Erlykin and Wolfendale \cite{erlykin97a} trying to explain 
possible structures around the knee, as well as several
particle physics motivated scenarios which try to explain the
knee due to different kinds of \cray-interactions.  For example,
photodisintegration at the source \cite{candia00}, interactions
with gravitationally bound massive neutrinos \cite{wigmans98}, or
sudden changes in the character of high-energy hadronic
interactions during the development of extensive air shower
(\eas) \cite{nikolsky95} have been considered.

To constrain the SN acceleration model from the other proposed
mechanisms precise measurements of the primary energy spectrum
and particularly of the mass composition as a function of energy
are needed.  Significant progress has been made on these problems
in recent years, but the situation is far from being clear.

The other target of great interest is the energy range around the
Greisen-Zatsepin-Kuzmin ({\sc gzk}) cut-off at $E \simeq 5\cdot
10^{19}$ eV. Explanation of these particles requires the
existence of extreme powerful sources within a distance of
approximately 100 Mpc.  Hot spots of radio galaxy lobes -- if
close enough -- or topological defects from early epochs of the
universe would be potential candidates.  This topic has been
addressed at this Symposium in some detail by Ostrowski,
Zavrtanik, and others.  Therefore, I will be very brief on
experimental aspects relevant to measurements in this energy
range.

\section{Extensive Air Showers and Experimental Observables}

An air shower is a cascade of particles generated by the
interaction of a single high energy primary cosmic ray particle
with the atmosphere.  The secondary particles produced in each
collision - in case of a primary hadron mostly charged and
neutral pions - may either decay or interact with another
nucleus, thereby multiplying the particles within an \eas.  After
reaching a maximum (\xm) in the number of secondary photons,
electrons, muons, and hadrons, the shower attenuates as more and
more particles fall below the threshold for further particle
production.  A disk of relativistic particles extended over an
area with a diameter of some tens of meters at $10^{14}$ eV to
several kilometre at $10^{20}$ eV can then be observed at ground. 
This magnifying effect of the earth atmosphere allows to
instrument only a very small portion of the \eas\ area and to
still reconstruct the major properties of the primary particles. 
It is a lucky coincidence that at the energy where direct
detection of \cray s rays becomes impractical, the resulting air
showers become big enough to be easily detectable at ground
level.

Due to the nature of the involved hadronic and electromagnetic
interactions and the different decay properties of particles, an
\eas\ has three components, electromagnetic, muonic, and
hadronic.  On average, a 1 PeV primary proton will produce
about $10^{6}$ particles at sea-level, 80\,\% of which are photons,
18\,\% electrons, 1.7\,\% muons, and about 0.3\,\% hadrons. 
Neutrinos will also be produced by weak decays of particles, but
they remain unseen by ordinary \eas\ experiments.  During their
propagation through the atmosphere, relativistic charged
particles will also produce Cherenkov light and, finally,
excitations of the $2P$ and $1N$-band of N$_{2}$ and N$_{2}^{+}$,
respectively, will give rise to emission of fluorescence light. 

Extracting the primary energy and mass from such measurements is
not straightforward and a model must be adopted to relate the
observed \eas\ parameters to the properties of the primary
particle.  As shall be discussed below, the analysis is
complicated by large fluctuations of \eas\ observables and by the
fact, that virtually all of the \eas\ observables are sensitive to
changes in the mass {\em and} energy of the primary particle so
that a careful reconstruction of the energy spectrum requires
knowledge of the chemical composition.

\begin{wrapfigure}[25]{l}{9.8cm}
\epsfig{file=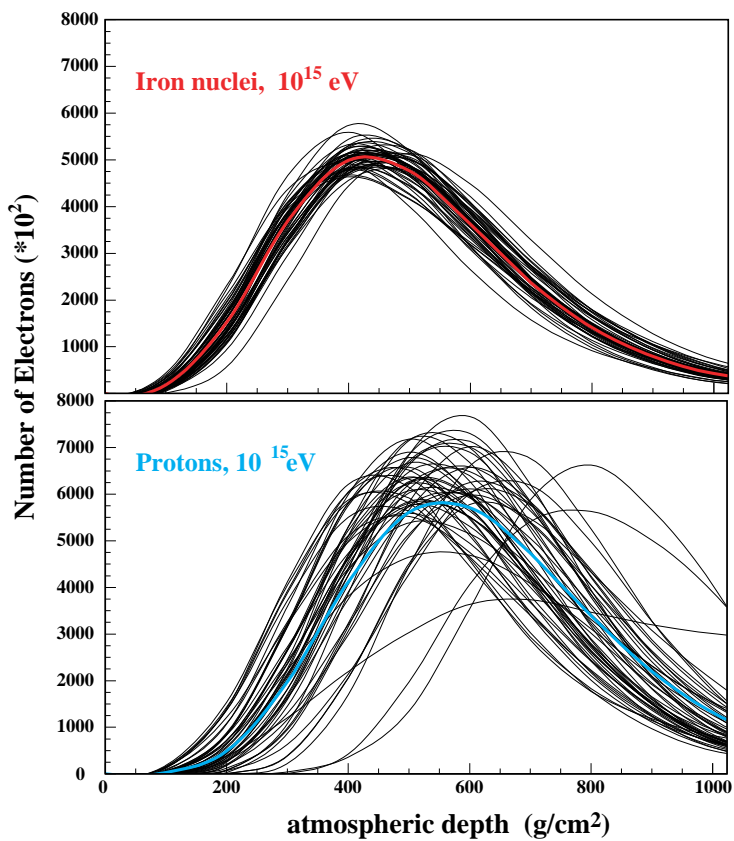,width=9.8cm}\vspace*{-2mm}
\caption{CORSIKA \cite{corsika} simulations of the longitudinal
shower development using the QGJSET \cite{qgsjet} hadronic
interaction model. The two thick lines represent averages of the 
individual p and Fe showers.}
\label{fig:long-dev}
\end{wrapfigure}

Ignoring these complications for a moment, some basic
characteristics of \eas, as observed in Fig.~\ref{fig:long-dev},
can be deduced from very simple considerations.  On average, the
first interaction of the primary particle with a nucleus in the
atmosphere - mostly nitrogen - will occur within its
characteristic interaction length, $\lambda_{I} \propto
1/\sigma$.  Due to the larger hadronic cross section of a Fe+N
system as compared to p+N, the iron nucleus will - on average -
interact higher in the atmosphere.  Because of the strong
absorption of the electromagnetic and hadronic shower components
in the atmosphere, one expects
to detect fewer electrons and hadrons at ground in case of
primary heavy nuclei.  In a simplified picture, we furtheron may
consider the primary iron nucleus as 56 nucleons each having
1/56$^{th}$ of the primary energy.  This has two important
consequences: i) the energy of produced particles within a single
collision will be lower by about the same number, so that more
pions and kaons have a chance to decay into muons before
reinteracting with another air nucleus, ii) the multiplicity of
produced particles within a single high energy collision roughly
scales with the logarithm of the energy, so that 56 nucleons each
with 1/56$^{th}$ of the energy, will produce a higher number of
secondary particles, which again may decay into muons.  As a
result of the two effects one expects to observe more muons at
ground in case of heavy primaries.  Also, fluctuations in the
number of particles will be smaller in case of heavy primary
nuclei.  Major experiments performing particle measurements at
ground in the energy range of the knee include CASA-MIA
\cite{borione94}, EAS-TOP \cite{eastop-99b}, HEGRA \cite
{arqueros00}, KASCADE \cite{kascade-90,kascade-97c}, and
Tibet-AS$\gamma$ \cite{tibet-96c}, only the latter two of which
currently take data.  Most of the data in the {\sc gzk} energy
range are based on the Akeno Giant-Air-Shower-Experiment (AGASA)
\cite{teshima92}.  Their primary observables are lateral particle
density distributions at ground, $\rho_{e,\mu,h}(r)$, and their
integrals $N_{e,\mu,h} = \int_{r_{1}}^{r_{2}} 2\pi r
\rho_{e,\mu,h}(r) dr$, yielding total (for extrapolations $r_{1}
\to 0$ and $r_{2} \to \infty$) or truncated particle shower
sizes. Truncated muon-sizes, $N_{\mu}^{\rm tr}$, have been 
introduced by the KASCADE collaboration in order to avoid 
systematic uncertainties resulting from extrapolations of 
$\rho_{\mu}(r)$ to large distances not covered by the 
experiment. 

\begin{table}[tbp]
    \centering
    \caption[xx]{Compilation of experimental \eas\ observables and 
    operating experiments. References are given in the 
    text.\\[.2ex]}
    \label{tab1}
\small
    \begin{tabular}{|l|l|c|l|}
    \hline
    Measurement of    & Observables & Energy Ranges & 
    Operating Expts.\  \\
    \hline
    charged particles & shower size & $\ga 5 \cdot 10^{13}$ eV &  
    GRAPES, \\
                      & lateral density distr. & & KASCADE, \\
		      & arrival time distr. & & Maket ANI, SPACE \\
    \hline
    muons             & muon size & $\ga 5 \cdot 10^{13}$ eV  & 
    KASCADE, \\
                      & lateral muon distr. & &  AGASA ($> 
                      10^{18}$ eV) \\
		      & arrival time distr. & & \\
		      & $\mu$-production height & $\ga 10^{15}$ eV  & \\ 
		      \hline
    hadrons           & hadron size & $5 \cdot 10^{13}$ - $ 10^{17}$ &
    KASCADE  \\
                      & lateral hadron distr. & (practical limit & \\
		      & energy distribution &  by detector  & \\
		      & spatial correlations &  area) & \\
    \hline
    Cherenkov light   & Cherenkov size & $5 \cdot 10^{13}$ - $ 
    10^{17}$ & BLANCA, HEGRA \\
                      & lateral Ch.\ distr. & (practical limit & 
                      DICE, TUNKA\\
		      & shower shape &  by area of Ch.-cone) & \\
    \hline
    Fluorescence light& total light yield & $\ga 10^{17}$ eV &  
    HIRes \\
                      & longitudinal profile & (limited by & \\
		      & time profile & signal/noise) & \\
    \hline
\end{tabular}
\normalsize
\vspace*{3mm}
\centering
\caption[xx]{Compilation of \eas\ observables and their 
sensitivity to primary energy and mass. ($^{(*)}$: energy 
estimates dependent also on primary mass)\\[.2ex]}
\label{tab2}
\begin{tabular}{ll}
\hline
Energy determination    & Mass determination  \\
\hline
$a \cdot \lg N_{\mu} + b \cdot \lg N_{e}$  &
            $\lg N_{e} / \lg N_{\mu}$ \\
  & shape of electron lateral distribution; age-parameter \\
  & mean muon production height \\
  & muon arrival time distributions \\
  & $\lg N_{h} / \lg N_{\mu}$, $\lg \sum E_{h} / \lg N_{\mu}$, 
  spatial distr.\ of hadrons \\
  & fluctuations of \eas\ parameters, e.g.\ $\lg (N_{\mu}) / \lg 
  (\overline{N}_{\mu})$ \\
$\rho_{Ch}(120)$ $^{(*)}$ & non-imaging counters: inner Cherenkov slope\\
total Ch.\ light $^{(*)}$ & Ch.\ telescopes: shape of shower image\\
		 & (the Ch.\ observables are often considered a measure of \xm) \\
total fluorescence light & position of shower maximum \\
\end{tabular}
\end{table}

Besides measuring particles at ground, experiments also aim at
observing traces of the longitudinal shower development.  A basic
parameter here is the position of the shower maximum, \xm, which
penetrates deeper into the atmosphere with increasing primary
energy and decreasing primary mass.  Experimentally, \xm\ is
often inferred from the lateral density distribution of Cherenkov
photons, $\rho_{Ch}$.  Experiments following this approach
include CASA-BLANCA \cite{fortson99} and the HEGRA AIROBICC
detectors \cite{karle95}.  Also, air Cherenkov telescopes have
been employed for sampling the shower maximum (HEGRA
\cite{arqueros00}, DICE \cite{kieda99}), and more recently, \xm\
has been reconstructed from muon tracks at ground by means of
triangulation (HEGRA \cite{bernloehr96} and KASCADE (talk
presented by C.\ B\"{u}ttner at this Symposium)).  A very elegant
approach is the detection of air fluorescence light at large
distances from the shower axis giving {\em true\/} information
about the longitudinal development of the \eas.  The major
drawback is the low light yield limiting its application to
energies $E \ga 10^{17}$ eV. This technique pioneered by Fly's
Eye \cite{bird94}, is being employed by HIRes \cite{abuzayyad00c}
and the Pierre Auger Observatory \cite{auger-www}.  A summary of
the different experimental observables is given in
table~\ref{tab1}.  Deep underground muon detectors are omitted
here because of limited space and because of their only indirect
relation to \eas\ observations.

\section{Techniques of Data Analysis and Selected Results}
\subsection{Single parameter methods}

Until recently, most experiments have performed single parameter
analyses using only one of the \eas\ observables discussed above. 
The composition has then been inferred by comparing the mean
value of the experimental observable to either different trial
composition models or to mean values of simulated proton and iron
primaries and interpolations to a hypothetical `mean mass'. 
Obviously, a significant piece of information available from the
full {\em distribution} of shower observables is ignored in such
analyses.  The problems here are manifold, most importantly: i)
the extracted mean mass is ambiguous, i.e.\ the mean values of
the experimental distribution can be described by a single `mean
mass' component, by a mix of pure protons and iron, or by any
other combination.  Clearly, a `composition' in a strict sense is
not obtained.  ii) There is no clue of how well the simulations
describe the experimental data.  Only if the mean value of the
experimental observable falls outside the window spanned by
\begin{wrapfigure}[20]{l}{9cm}
\vspace*{-5mm}\epsfig{file=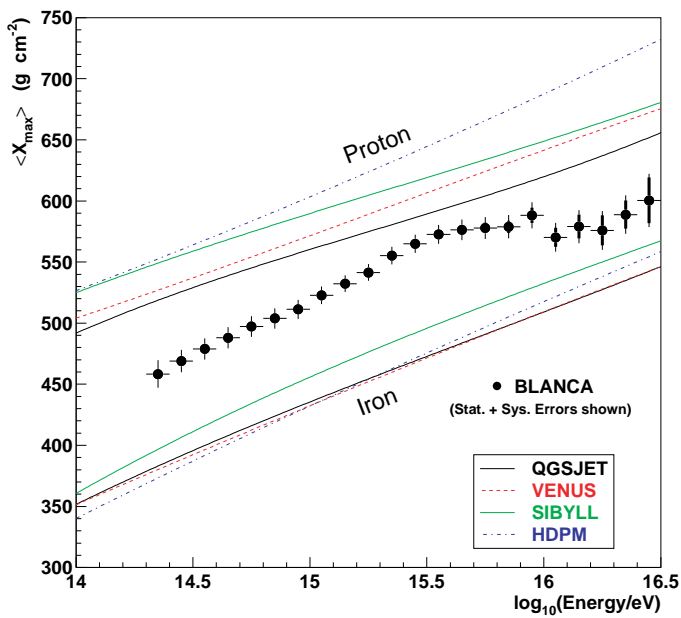,width=9cm}\vspace*{-3mm}
\caption{The mean depth of shower maximum \xm\ as a function
of energy deduced from CASA-BLANCA \cite{fowler-2000}.
The lines represent the values for pure 
proton and pure iron samples as predicted by different hadronic
interaction models.}
\label{fig:xmax-blanca}
\end{wrapfigure} 
proton and iron simulations, doubts can be raised about the
quality of the simulations (or apparatus).  Furthermore,
publications discussing single parameter results often do another
simplification by comparing measured and simulated observables
not at the level of the detector, but instead comparing quantities
which are {\it inferred\/} from experimental data by means of \eas\
simulations.  A very prominent example of such kind of analyses
are plots of \xm\ as a function of energy as shown in
Fig.~\ref{fig:xmax-blanca}.  In this and many
other cases, \xm\ is based on measurements of the slope of the
Cherenkov lateral distribution within about 120 m from the
shower core.  \eas\ simulations using different primary particles
are then performed to convert this experimental slope parameter
to \xm.  Finally, this biased and model dependent \xm-value is
presented in such graphs and compared to \xm-values obtained
directly from CORSIKA simulations.  Clearly, comparing
experimental and simulated slope parameters directly instead of
introducing secondary quantities appears much more appropriate and
avoids sources of systematic biases.  Determination of $E_{prim}$
in experiments using open Cherenkov counters is mostly based
on the Cherenkov light intensity observed at about 120 m from
the shower core where influences of the unknown primary mass are
minimal.  However, a careful analysis of this procedure exhibits
a residual mass dependence on the order of $\pm 10$\,\% in the
knee energy range \cite{pryke-2000}.  Also,
uncertainties to the energy scale by at least the same amount are
caused by hadronic interaction models \cite{fowler-2000} but,
again, are not considered in plots of the type of
Fig.\,\ref{fig:xmax-blanca}.  Finally, the figure also nicely
demonstrates the aforementioned inabilities to judge the quality
of the different hadronic interactions models or to infer a true
composition.  Despite this criticism, the qualitative feature of
the data -- a somewhat lighter `mean mass' towards the knee and a
heavier one above -- appears common to all interaction models.

\begin{figure}[t]
\begin{center}
\epsfig{file=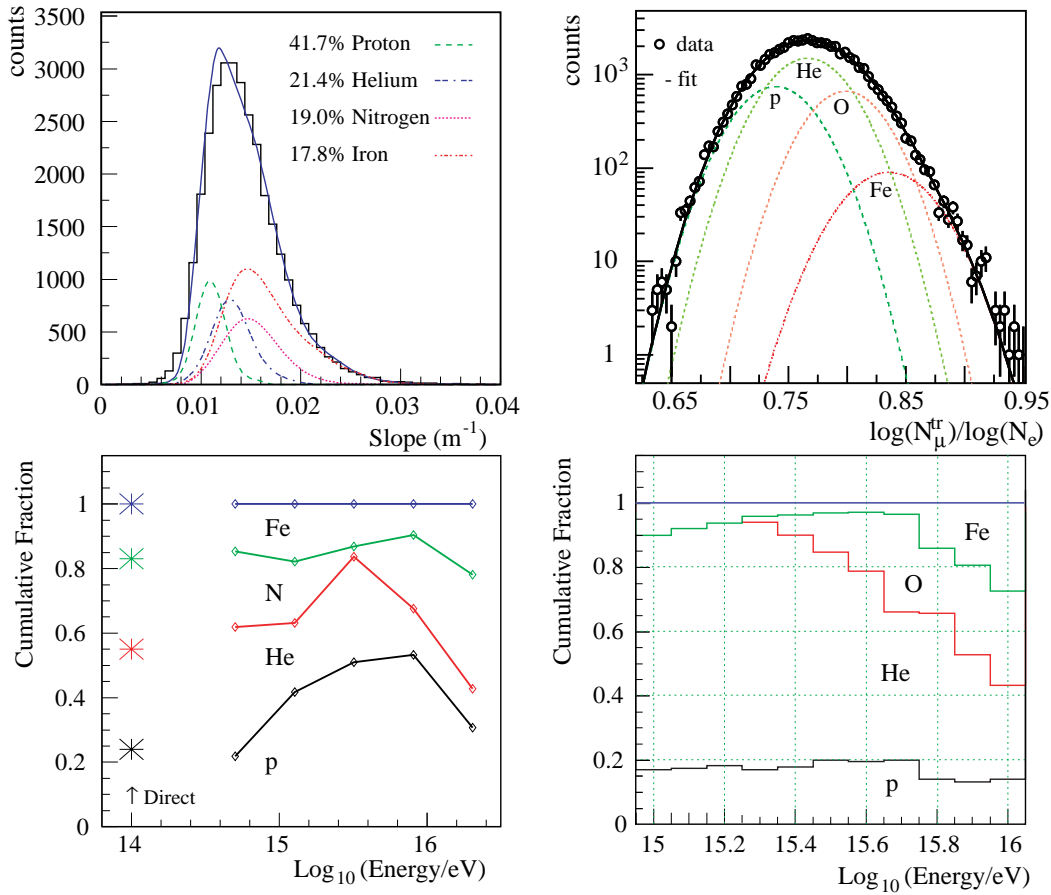,width=14cm}
\end{center}
\caption[xx]{Event-by-event distributions of the inner Cherenkov
slope (l.h.s.) \cite{fowler-2000} and of the muon/electron ratio
(r.h.s.) \cite{weber99}.  The lower panel shows the cumulative
fractions of different primary particles as a function of their
energy.}
\label{fig:shapes}
\end{figure}

With increasing computer power, \eas\ simulations have advanced
in quality and quantity, i.e.\ both the complexity to which
details of hadronic interactions and propagation effects in the
atmosphere are taken into account has been improved as well as
the number of simulated events available for comparison. 
Therefore, analyses of \eas\ data could progress from comparing
only mean values to full distributions of observables, the
benefits of which have been discussed already.  Two examples are
presented in Fig.\,\ref{fig:shapes}.  The left hand side of the
top panel presents the event-by-event distribution of the inner
Cherenkov-slope (previously used as a measure for \xm, see
above) in an energy range $8\cdot10^{14} \le E \le
2\cdot10^{15}$~eV \cite{fowler-2000} and the right hand side the
distribution of $\log N_{\mu}^{\rm tr}/\log N_{e}$ obtained from KASCADE
for zenith angles $18^{\circ}$ - $25^{\circ}$ and similar
energies \cite{weber99}.  Both data sets are compared to CORSIKA
simulations of different primary particles using the QGSJET
model.  Note, that the left hand tails of the distributions are
well accounted for by proton simulations while the right hand
tails are accounted for by iron simulations.  The KASCADE
distributions cover more than three orders of magnitude and thus
provide some level of confidence in the results of the
simulations.  Both data sets also require at least 3 different
primary masses to account for the full shape of the experimental
distributions.  However, it is also observed that the
Cherenkov data alone do not provide sufficient discrimination
power to distinguish the nitrogen from the iron distribution. 
The bottom panel shows the extracted mass fractions as a function
of energy.  Clearly, a heavier composition is observed at
energies above the knee.  Also, the relative abundance of iron is
fairly similar in both data sets but the major difference is a
reversed proton and Helium abundance resulting in a somewhat
lighter composition from KASCADE data.  To resolve this
differences, several cross checks of the two data sets appear
expedient, e.g.\ tests of the zenith angle dependence, still
higher statistics in simulations to verify the tails of the
distributions, etc. Such work is in progress now.

\subsection{Multi-component methods and non-parametric approaches}

Multi-component detector installations can measure several \eas\
parameters simultaneously on an event-by-event basis.  Proper
combinations can then be identified to provide an estimate of the
primary energy and mass (see Table 1,2).  The analysis of
multivariate parameter distributions also needs to account for
influences of fluctuations which may be different in different
observables.  Non-parametric Bayesian methods and neural network
approaches are well suited for these purposes and they also
specify the uncertainty of the results in a quantitative way
\cite{chili89}.  Simpler approaches, like the {\sc knn}-method
have also been employed \cite{glasmacher99b}, but they suffer
from various deficiencies like dependencies of the results on the
density and thereby on the statistics of simulated data points. 

First comprehensive non-parametric approaches with promising
results have been presented at this Symposium by M.\ Roth {\it et
al\/}.  In these techniques, each event is represented by an
observation vector $x = ( N_{\mu}^{\rm tr}$, $N_{e}$, $N_{h}
\ldots )^{T}$ of $n$ suitable \eas\ observables.  This vector
serves as input to a procedure based on Bayesian or neural
network decision rules by which the observed event is assigned to
a given class of elemental groups, say p, O, and Fe.  The first
step in such an analysis is a calculation of likelihood
(probability density) distributions $p(x|\omega_{i})$ by means of
Monte Carlo simulations using different primary particles.  Here,
$p(x|\omega_{i})$ describes the probability to observe the
multi-dimensional observation vector $x$ in an event belonging to
a given class $\omega_{i} \in \{$p, O, Fe$\}$.  The term
`non-parametric' indicates that the representations of the
distributions (like probability density functions of Bayes
classifiers or weights of neural networks) are no more specified
by a-priori chosen functional forms, but are constructed through
the analysis process by the given (simulated) data distributions
themselves.  Using the Bayes theorem one can then translate the
likelihood $p(x|\omega_{i})$ for finding an event $x$ in a given
class $\omega_{i}$ to the actually required distribution
$p(\omega_{i}|x)$ representing the probability of class
$\omega_{i}$ being associated to a measured event $x$.  In this
step, an assumption is made about an a priori knowledge of the
relative abundance of each class, a subject being handled by
Bayesian inference procedures.  If there is no further knowledge,
the prior probabilities should be assumed to be equal (Bayes'
Postulate).

Analysing an experimentally measured event, a statistical
decision on the primary particle type/energy is then to be made. 
The applied Bayesian approach of the statistical inference
provides the {\em Bayes optimal\/} way of combining prior and
experimental knowledge and the Bayes theorem specifies how such
modification should be made \cite{Raifa61}.  In simplest case,
the Bayes optimal decision rule is to classify $x$ into class
$\omega_i$, if $p(\omega_i|x)>p(\omega_j|x)$ for all classes
$\omega_j\neq \omega_i$.  In this way one is also able to specify
the actual quality of the decision by classification and
misclassification matrices of the results.  Unfortunately,
because of limited computer time, analyses making use of all
these pieces of important information have not yet been performed
for high statistics data.  However, good agreement to
faster neural network approaches was reported in Ref.\
\cite{roth-nonpara00}.

It is important to realize that the composition extracted that
way yields a somewhat heavier composition as compared to the
results from Fig.\,\ref{fig:shapes}.  This effect may possibly
be caused by different treatments of correlations among the
fluctuating \eas\ observables.  Furthermore, detailed
investigations also show that within a given interaction model,
different sets of \eas\ observables lead to different mass
compositions \cite{roth-nonpara00}.  Generally, inclusion of
hadronic observables tends to shift the composition to a heavier
one, a result already suggested by earlier studies of the KASCADE
collaboration \cite{kascade-99c}.  These inconsistencies appear
to be caused by deficiencies of the employed hadronic interaction
models.  Both kinds of effects are subject to further studies and,
at present, any {\em quantitative\/} result on chemical
composition may be subject to further changes.

Results on chemical composition are often presented in terms of
the mean logarithmic mass.  However, this quantity has some
severe drawbacks as any information about individual abundances
is lost.  Furthermore, in absence of an unique prescription on
how to calculate the average from abundances of different mass
groups, uncertainties in $\langle \ln A \rangle$ may be on the
order of 0.2 and more.  This has been source for some confusion
in the past and can best be avoided by providing tables of
reconstructed abundances of
\begin{wrapfigure}[18]{l}{10cm}
\epsfig{file=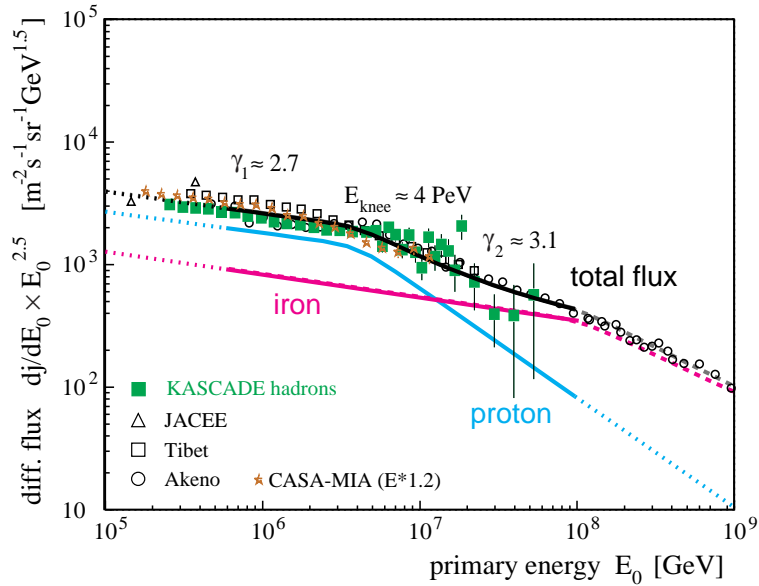,width=10cm}
\caption{Primary energy spectrum as obtained from different 
experiments \cite{glasstetter99}. (See text for references.)}
\label{fig:e-spec}
\end{wrapfigure} 
mass groups (including correlated
uncertainties).

A compilation of the all-particle energy spectrum
reconstructed from various experiments is presented in figure
\ref{fig:e-spec}.  The agreement appears reasonable and
deviations are mostly explained by uncertainties in the energy
scale by up to 25\,\%, e.g.\ CASA MIA data \cite{glasmacher99a}
were shifted upwards in energy by 20\,\% to yield a better
agreement to the other data sets.  This is likely to be explained
by the interaction model SIBYLL 1.6 \cite{sibyll16} employed by
the authors of Ref.\,\cite{glasmacher99a} but which has been
proven to provide only a poor description of the experimental
data \cite{kascade-99c}.  The lines correspond to a simultaneous
fit of the electron and muon size spectra of KASCADE, assuming
the all-particle spectrum to be described by a sum of proton and
iron primaries \cite{glasstetter99}.  Interestingly, a knee is
only reconstructed for the light component and no indication of a
break is seen in the heavy component up to about $10^{17}$~eV.
This important finding giving direct support to the picture of
acceleration in magnetic fields (see above) will be target of
future studies with improved experimental capabilities.  For
example, KASCADE and EAS-TOP have just started a common effort to
install the EAS-TOP scintillators at the site of
Forschungszentrum Karlsruhe providing a 12 times larger
acceptance as compared to the original KASCADE experiment and
still taking advantage of the multi-detector capabilities.

\section{Summary and Outlook}

Much progress has been made in reconstructing the parameters of
primary \cray-particles from \eas\ observables.  The advent of
multiparameter measurements and realistic \eas\ simulations now
allow for much more detailed investigations taking account for
\eas\ fluctuations and detector effects.  Different experiments
agree fairly well on the all-particle energy spectrum but there
is still some uncertainty in extracting the chemical composition. 
This may partly be explained by uncertainties of hadronic
interaction models employed in \eas\ simulations and there
appears a demand for more tests of hadronic interaction models by
means of accelerator and/or \eas\ data.  Other sources of
systematic uncertainties appear to be caused by incomplete data
analysis techniques and biases affecting mostly single parameter
measurements.  As discussions and presentations at this Symposium
have shown, there is well founded hope to solve these
imperfections thereby answering the question about the sources of
\cray s and the origin of the knee already in the very near future.

\vspace*{-5mm}
\paragraph{Acknowledgement}
{\small I would like to express my gratitude to the organizing
committee for their hospitality and inviting me to give this
talk.  During the preparation I greatly benefited from fruitful
discussions with my colleagues from the KASCADE collaboration,
with C.\ Pryke, J.\ Matthews, S. Swordy, and many others. I would 
also like to thank M. Giller for valuable comments to this
manuscript.}

\section*{References}

\end{document}